\documentstyle[prd,aps,preprint]{revtex}
\begin{document}
\draft

%
%

\preprint{Nisho-2003-3}
\title{Color Ferromagnetism 
and Quantum Hall states in Quark Matter}
\author{Aiichi Iwazaki}
\address{Department of Physics, Nishogakusha University, Shonan Ohi Chiba
  277-8585,\ Japan.} 
\date{Nov 1, 2003} \maketitle
\begin{abstract}
We discuss a possibility of the presence of 
a stable color ferromagnetic state
in SU(2) gauge theory of quark matter; 
a color magnetic field is spontaneously generated due to
the gluon's dynamics. The state arises between the hadronic state and 
the color superconducting state when the density of quarks is varied.
Although the state has been known to have unstable modes,
we show that unstable modes form quantum Hall states,
in which the instability disappears.
Namely, the quark matter possesses a stable phase with the ferromagnetic state
and the quantum Hall state of gluons.  
\end{abstract}
\hspace*{0.3cm}
\pacs{PACS 12.38.-t, 12.38.MN, 24.85.+p, 73.43.-f  \\ 
Quark Matter,\,\,Color Superconductivity,\,\, Quantum Hall States
\hspace*{1cm}}
\tightenlines
Quark matter is expected to possess several phase structures when its temperature and
densiy are varied; the hadronic phase, the quark gluon plasma phase, and 
the color superconducting phase. 
The superconducting phase\cite{colors} has attracted much attention recently.
Although the observation of the phase in present experiments of heavy ion collisions
is difficult, the phase might be observed in compact stars such neutron stars or quark stars.

In this report we point out a possible existence of a color
ferromagnetic state\cite{iwa,savidy} in which the color magnetic field $\vec{B}$ is
spontaneously generated. The phase can be observed easier in
the experiments than the superconducting phase can be. 
We show that the phase is realized due to the
dynamical effects of gluons, although quarks play an important role for the 
realization. 
It is very intriguing that a quantum Hall state\cite{qh} of some gluons	 
is formed to stabilize the ferromagnetic state. 
The gluons has been known previously to be unstable
modes\cite{savidy,nielsen} in the color magnetic field.
Such a quantum Hall state is shown to
carry color charges. The charges must be supplied by the quark matter
because we are considering color neutral system.
Thus, the phase is realized only when the quark matter is present. 


We consider the SU(2) gauge theory 
with massless quarks of two flavours.
As is well known\cite{savidy,nielsen}, the effective potential $V$ of the constant color magnetic 
field has the minimum at non-zero magnetic field in one loop
approximation:
$V= \frac{11}{48\pi^2}e^2B^2\left(\log
  (eB/\Lambda^2)-\frac{1}{2}\right)-\frac{i}{8\pi}e^2B^2$,
with an appropriate renormarization of the gauge coupling $e$,
where directions of the magnetic field in real space and color space
are arbitrary. In any case of their choices the spontaneous generation of
the magnetic field breaks the spatial rotational symmetry and the color gauge
symmetry. The SU(2) gauge symmetry is broken into the U(1) gauge symmety. 
This non-trivial minimum apparently implies the spontaneous generation
of the color magnetic field. But, 
the presence of the imaginary part of the potential 
implies that a state with $B\neq 0$, but with no
condensation of some gauge field is unstable. Namely it 
leads to
excitation of unstable modes around the minimum $eB_{mim}=\Lambda^2$.
This is similar to 
the case that when we expand a potential $=a^2(|\phi|^2-v^2)^2/4$ of a scalar field
around the local maximum, $\phi=0$, i.e. wrong vacuum, unstable modes
are present. We know that
their excitations lead to the stable vaccum $\langle \phi \rangle=v$ with condensation of a spatially uniform
unstable mode.
In the gauge theory, similar condensation of gluon's unstable modes in
the magnetic field
is expected to arise. 


In order to explain the unstable modes in the gauge theory,
we decompose the gauge fields $A_{\mu}^i$ such that
$
A_{\mu}=A_{\mu}^3,\,\,\mbox{and} \,\,
\Phi_{\mu}=(A_{\mu}^1+iA_{\mu}^2)/\sqrt{2}$ 
where indices $1\sim 3$ denote color components. Then, we may suppose 
that the field $A_{\mu}$ is the 'electromagnetic' field 
of the U(1) gauge symmetry and $\Phi_{\mu}$ is the charged vector field,

\begin{eqnarray}
\label{L}
L&=&-\frac{1}{4}\vec{F}_{\mu
  \nu}^2=-\frac{1}{4}(\partial_{\mu}A_{\nu}-\partial_{\nu}A_{\mu})^2-
\frac{1}{2}|D_{\mu}\Phi_{\nu}-D_{\nu}\Phi_{\mu}|^2- \nonumber \\
&-&ie(\partial_{\mu}A_{\nu}-\partial_{\nu}A_{\mu})\Phi_{\mu}^{\dagger}\Phi_{\nu}+\frac{e^2}{4}(\Phi_{\mu}\Phi_{\nu}^{\dagger}-
\Phi_{\nu}\Phi_{\mu}^{\dagger})^2
\end{eqnarray}
with $D_{\mu}=\partial_{\mu}+ieA_{\mu}$,
where we have omitted a gauge term $D_{\mu}\Phi_{\mu}=0$. 
Using the Lagangian we can show that the energy $E$ of
the charged vector field $\Phi_{\mu}\propto e^{iEt}$ 
under the magnetic field $\vec{B}$ is given by
$E^2=k_3^2+2eB(n+1/2)\pm 2eB$  
with a gauge choice, $A^B_{j}=(0,x_1 B,0)$ and $(\partial_{\mu}+ieA^B_{\mu})\Phi_{\mu}=0$, 
 where we have taken the spatial direction of $\vec{B}$ being along $x_3$ axis.
$\pm 2eB$ ( the integer $n\geq 0$ ) denote contributions of spin components
of $\Phi_{\mu}$ ( Landau levels )
and $k_3$ denotes
momentum in the direction parallel to the magnetic field.
We should note that in each Landau level, there are many degenerate
states specified by momentum $k_2$, whose degeneracy is given by 
$eB/2\pi$ per unit area; for example, wave functions of the lowest
Landau level ( $n=0$ ) is given by
$\Phi_{\mu} \sim e^{-ik_2x_2-ik_3x_3}\exp(-eB(x_1-k_2/eB)^2/4)$ where
the energy, $E=\sqrt{k_3^2+eB \pm 2eB}$, is degenarate in $k_2$.

We find that the states with parallel magnetic moment ($-2eB$) 
in the lowest Landau level ( $n=0$ )
are unstable when $k_3^2<eB$; they are unstable modes.
Therefore, we expect from the lesson
in the scalar field
that the spatially uniform unstable modes ( $k_3=0$ ) are
excited to lead to a true stable vacuum with their condensation. 
Actualy, there have been several
attempts\cite{nn} 
to find the true vacuum by making the condensation of the unstable modes. 
But, it was difficult to see
whether or not any unstable modes dispappear in the condesed state.
Here we should note that the unstable modes with $k_3=0$, which must
condense, occupy the lowest Landau level and are spatially two
dimensional bosons. 
The situation is quite similar to the one of 
the two dimensional electrons forming QHSs. 
The only difference is that 
gluons are bosons, while electrons are fermions.
In order to find the stable state in the gauge theory,
we extract only the unstable modes from the Lagrangian eq(\ref{L}),
ignoring the other modes coupled with the unstable modes and 
obtain two dimensional Lagrangian,

\begin{equation}
L_{unstable}=|(i\partial_{\nu}-eA_{\nu}^B)\phi_u|^2+2eB|\phi_u|^2-\frac{\lambda}{2} |\phi_u|^4, 
\end{equation}
with $\lambda=e^2/l$, where the field
$\phi_u=(\Phi_1-i\Phi_2)\sqrt{l/2}$
denotes the unstable modes in the lowest Landau level.
$l$ denotes the coherent
length of the magnetic field, namely, 
its extention in the direction of the field and  
the index $\nu$ runs from $0$ to $2$.
The presence of the negative mass term implies
that the state $<\phi_u>=0$ is unstable.

This Lagrangian is quite similar
to a Lagrangian in the Landau Ginzberg theory of the superconductivity.
It apparently seems that the groundstate is simply given by
$<\phi_u>= \sqrt{2eB/\lambda}$, the condensed states of the field $\phi_u$. 
But it is impossible because the term of $A_{\mu}^B$ is present in the
kinetic term. If this term vanishes, the term of the negative mass
also vanishes so that the solution $<\phi_u>\neq 0$ does not
exist. Physically, the Lagrangian $L_{unstable}$ describes such a system that 
the particles of $\phi_u$ move in the magnetic field and interact with
each other through a delta function repulsive potential.
We know from numerical simulations that the nonrelativistic particles 
with such an interacting potential
can form a Laughlin state even if they are bosons;
the Laughlin state is an explicit form of the wave function representing a qantum Hall state. 

In order to see explicitely the QHS of the field,
we introduce Chern-Simons gauge field $a_{\mu}$ to make composite gluons;
bosons attached with the Chern-Simons flux. Then, a rellevant
Lagrangian\cite{zhang,qh} is given by 

\begin{equation}
\label{la}
L_a=|(i\partial_{\nu}-eA_{\nu}+a_{\nu})\phi_a|^2+2eB|\phi_a|^2-\frac{\lambda}{2}|\phi_a|^4-
\frac{\epsilon^{\mu\nu\lambda}}{4\alpha}a_{\mu}\partial_{\nu}a_{\lambda},
\end{equation}
with antisymmetric tensor $\epsilon_{\mu\nu\lambda}$ 
with $\epsilon^{012}=\epsilon_{12}$,
where the statistical factor $\alpha$
should be taken as $\alpha=2\pi\times $integer to keep the
equivalence of the system described by $L_a$ to that of $L_{unstable}$.
The equivalence 
has been shown \cite{seme} in a operator formalism although the
equivalence had been known in the path integral formalism using the world
lines of the $\phi_a$ particles.

This idea of composite gluons is very popular in the discussion of QHS
of electrons, where the idea of composite electrons is used. Namely,
electron field is replaced by composite electron field,
which is a boson field attached with Chern-Simons flux.
That is, particles with Fermi statistics can be described in two
dimensional space by
bosonic particles attached with a ficticious flux $2\alpha$ of Chern-Simons
gauge field $a_i$. Owing to this flux, the exchange of the bosonic
particles induces a phase $e^{i\alpha}$ in their wavefunction. Thus,
with the choice of $\alpha=\pi\times $odd integer,
the wavefunctions describe particles with Fermi statistics. 
Similarly gluons are described by composite gluons with the choice
of $\alpha=\pi\times $even integer.

QHSs can be described by the 
Chern-Simons gauge theory even in the mean field approximation \cite{qh}.
Equations of motion are given by

\begin{eqnarray}
\label{eq1}
\phi_a^{\dagger}\,i\partial_0\,\phi_a+c.\,c.+2a_0\,|\phi_a|^2=\frac{1}{4\pi}\epsilon_{ij}\,\partial_i\,a_j
\\
\phi_a^{\dagger}\,(\,i\partial_i-eA_i+a_i)\,\phi_a+c.\,c.=\frac{1}{4\pi}\epsilon_{ij}\partial_0\,a_j 
\\
(i\,\partial_0+a_0)^2\,\phi_a-(\,i\vec{\partial}-e\vec{A}+\vec{a}\,)^2\,\phi_a+2eB\,\phi_a=\lambda |\phi_a|^2\phi_a.
\end{eqnarray}
where we have taken $\alpha=2\pi $.

Using these equations, we can easily find a solution representing a quantum Hall state.
The essence is that in the QHS the ficticious flux is
canceled on average with the real magnetic flux, i.e. $a_i=A_i$.  
Consequently, the field $\phi_a$ does not feel any gauge field.
Then, the Lagrangian eq(\ref{la}) is reduced effectively to one
representing an usual double well potential.
Hence we find a uniform solution, 
$<\phi_a>=v $: $v$ is obtained by solving the above
equations which are reduced to 
$2a_0\,v^2=\frac{1}{4\pi}\epsilon_{ij}\,\partial_i\,a_j=eB/4\pi$
and $a_0^2+2eB=\lambda v^2$.
This state 
is just a QHS of the field $\phi_a$:
We can show\cite{zhang} that the Hall conductivity of the state is
given by $e^2/2\alpha$.
This state is stable against small fluctuations
around the state. Furthermore, we find that vortex excitations
arise on the condensed state $<\phi_a>\neq 0$ with their energies
being positive. Namely, the excitations are stable modes in
the quantum Hall state. Therefore, the newly formed condensted state of the unstable modes
is stable. In this way we find that the ferromagnetic state 
is stabilized by the formation of the gluon's QHS. 

We should point out that since
the condensate of $\phi_a$ possesses a color charge ( $\rho=eB/4\pi$ ),
the charge must be supplied from somewhere in color neutral system. 
Without the supplier of the color charge, the condensation
can not arise so that the quantum Hall state is not realized.
Thus, the color ferromagnetic state is not stabilized. 
Quark matter is a supplier of the color charge.
Hence, the stable ferromagnetic state
is possible only in dense quark matter;
some color charges of quarks are transmitted to 
the condensate. Then, the color charge density of the quark matter
with its radius $L$ 
should be larger than that of the color condensate, $\rho/L=eB/4\pi
L$. Thus, it follows that the chemical potential, 
$\mu$, should be larger than $(3\pi eB/4L)^{1/3}\sim 180 \,
\mbox{MeV}(eB/0.04\,\mbox{GeV}^2)^{1/3}(3\,\mbox{fm}/L)^{1/3}$ in order for the ferromagnetic 
phase to arise in the quark matter.
Since it is a value nessesary for the realization of the phase,
a critical value separating the two phases,
hadronic phase and ferromagnetic phase is larger than it.   

Now, we wish to discuss an observable effect of the ferromagnetic phase in quark matter
produced by heavy ion collisions.
Namely, an observable QED's magnetic field 
is produced by quarks rotating around
the color magnetic field.
If the number of the color positive charged quarks with a flavour ( for example, up quark )
is the same as that of the color negative charged quarks with the flavour,
the total real magnetic moment produced by the quarks vanishes.
But the number of the color positive charged quarks
and that of the color negative charged quarks is different in
the color neutral system due
to the gluon condensation with the color charges in the QHS.
Namely, some of color positive charges of the quarks are transmitted to the condensate.
Thus, the number of the quarks with color positive charge
is smaller than the number of the quarks with color negative charge.
Therefore, the real magnetic moment produced by, for example, up quarks 
does not vanish. Since the lump of heavy ions is positively charged,
this mechanism works; real total magnetic moment produced by 
the quarks with each flavour does not vanish. 
Taking $eB$ being several $0.01 \,\mbox{GeV}^2$ and
the radius of quark matter $L$ being several fm, 
we can show that the real magnetic field with strength $10^{14}\sim 10^{15}$ Gauss is
produced in the color ferromagnetic phase of 
the quark matter, 
which may be generated by heavy ion collisions.

Finally we should mention that the ferromagnetic phase
involving quantum Hall state of gluons 
arises even in SU(3) gauge theory.
In the SU(3) gauge theory, there are three types of unstable modes\cite{su3}
which have different color charges corresponding to $\lambda_3$ and $\lambda_8$,
maximal Abelian sub-algebra. Then, several types of quantum Hall states are
present. In any quantum Hall states the ferromagnetic state is stable.
Observable effects mentioned above are also present. 
Therefore, the quark matter produced by heavy ion collisions or present in
compact stars shows an intrincic phenomena specific to the phase, which is a useful
indication of the presence of the ferromagnetic phase.


\end{document}